\documentclass[twocolumn,twoside]{revtex4}
\usepackage{graphicx}
\usepackage{fancyhdr}
\begin{document}

\title{Future Neutrino Facilities}

%

\author{D. A. Harris}
\affiliation{Fermi National Accelerator Laboratory, Batavia, Illinois USA and \\
York University, Department of Physics and Astronomy, Toronto, Ontario, Canada}

\begin{abstract}
The fact that neutrinos have mass and can oscillate from one flavor to another has opened up a wide range of neutrino flavor measurements.   Those measurements could uncover the source of CP violation that lead to the baryon asymmetry present in the Universe today, and will also enable us to understand more about how the masses of the fundamental particles are generated.  This report describes the DUNE, Hyper-Kamiokande, and PINGU experiments, which each employ unique strategies to learn more about neutrino flavor.  

\end{abstract}

\maketitle

\thispagestyle{fancy}


\section{Introduction}

The fact that neutrinos have mass and can oscillate from one flavor to another has opened up a wide range of neutrino flavor measurements.  Similar to the quark sector, there is a mixing matrix that translates between the weak (flavor) neutrino eigenstates and the mass eigenstates.  However, because neutrinos interact so rarely, the field of neutrino physics is far from being able to achieve the level of precision that the quark sector has been able to achieve.   Nevertheless, it is important to probe this sector in detail, since it is a clear place where the standard model is broken.  Studying neutrino flavor physics may also uncover the source of CP violation in the early universe that lead to the large baryon asymmetry in the universe today.  

There are at least as many ways to study neutrino flavor sector as there are sources of neutrinos.  This report on future neutrino facilities will focus on measurements of neutrino oscillations made at hundreds of MeV to dozens of GeV and over hundreds to thousands of kilometers.  Neutrinos from accelerators and from the atmosphere provide this energy range, and by studying neutrinos of these energies over long distances, experiments are able to address both the mass ordering of the neutrinos, as well as to search for whether or not neutrinos violate Charge-Parity (CP) conservation.  

The three future facilities described here demonstrate three different strategies to study neutrino flavor.  The PINGU experiment, an extension of the DEEPCORE part of the ICECUBE experiment, will study neutrinos produced in the atmosphere and can study them between baselines of 80~km to 10,000~km.  The Hyper-Kamiokande experiment, which is an extension of the T2K experiment, will study neutrinos of several hundred~MeV produced by the J-PARC neutrino beam facility in Japan that have traveled a baseline of about 300~km.  The DUNE experiment, which uses a new detector and neutrino beamline in the United States, will study neutrinos over energies from 1 to 4~GeV that have traveled 1300~km.  The primary goal of PINGU is to measure the neutrino mass ordering, the primary goal of Hyper-Kamiokande is to measure CP-violation, and DUNE's goals include both a measurement of the mass ordering and of CP-violation.  

\section{Neutrino Mixing Matrix}

Neutrino mixing occurs because the mass eigenstates are not the same as the weak eigenstates, similar to mixing in the quark sector.  However, the matrix that describes neutrino flavor mixing differs drastically from the matrix that describes quark mixing, because the off-diagonal elements are large compared to the diagonal elements.  In addition, because the mass differences are so tiny, the oscillations for $~$GeV particles take hundreds of kilometers to develop, rather than the micron distance scales of flavor mixing measured at colliders.  

The two independent mass differences between the three (standard) mass eigenstates differ by a factor of thirty, so although some oscillation probabilities are large, the second order effects where CP violation occur are seen only in small differences between neutrino and antineutrino oscillation probabilities.  This informs the strategies of oscillation experiments looking for CP violation:  they must make precise measurements of few per cent oscillation probabilities, rather than simply see oscillations in the first place.  In other words, oscillation experiments are challenged not only with getting high statistics, they must also be careful to keep systematic uncertainties to the per cent level.  Because these oscillation probabilities must be determined as a function of time that the neutrino has had to evolve, a measure of the neutrino energy, the distance the neutrino traveled from point of creation to detection, and final state neutrino flavor are key requirements to measuring oscillations.  How well those three quantities are measured determines the systematic precision of the measurement.  

By studying neutrino oscillations one can determine whether there is CP violation in the lepton sector that might give rise to the baryon asymmetry.  In addition, by studying neutrino oscillations using neutrinos that travel over long distances through the earth, one can determine whether the neutrino mass spectrum is similar to that of the charged fermion mass spectrum (one light mass eigenstate) or inverted from that of the charged fermion mass spectrum (two light mass eigenstates).  This is because the presence of electrons in the earth change the potential that the electron neutrinos and antineutrinos see while propagating, but does not affect the muon- or tau-flavored neutrinos\cite{wolf}. 

There are several neutrino oscillation experiments currently running:  two accelerator-based neutrino experiments, T2K and NOvA, use narrow band neutrino beams to study electron neutrino (antineutrino) appearance from a beam of muon neutrinos (antineutrinos).  These experiments have been running for one half to one decade, and in order to reach precision measurements of the CP violating phase if it is in fact non-zero, they would have to run for extremely long times.  A new generation of accelerator-based experiments is being designed with improved proton sources for more intense neutrino beams, larger detectors, and improved systematic uncertainties coming from better near detector facilities to measure neutrino interactions.  

\section{Neutrino Sources}

The first compelling evidence for neutrinos changing flavors over terrestrial distance scales came from measurements of the up-down asymmetry of events coming from neutrinos produced in the atmosphere\cite{kamiokande-atm}.  When high energy protons in cosmic rays strike the upper atmosphere, they produce pions and kaons which then decay on their way to the surface of the earth.  For every positively pion that decays, to first order one muon neutrino and one anti-muon are produced.  Then if that anti-muon decays on its way to the earth's surface, it will produce a positron and an electron neutrino, as well as a muon antineutrino.  This process results in a muon neutrino and a muon antineutrino for every electron neutrino, a ratio that is roughly insensitive to the initial cosmic ray energy spectrum, and to first order insensitive to the pion production cross section.  Of course high energy protons do not only produce positive pions, they also produce negative pions, and kaons of both charges.  Although uncertainty remains in the absolute neutrino flux prediction for each flavor and helicity of neutrino, the ratio of neutrino to antineutrino, and electron to muon flavor is relatively robust.  Atmospheric neutrino experiments take advantage of the fact that the zenith angle of the outgoing lepton from the neutrino interaction is on average the zenith angle of the incoming neutrino.   This in turn gives an approximate measure of the distance the neutrino has traveled from where it has been produced, which determines the time the neutrino has had to change flavors.  Neutrinos that produce leptons that propagate from the top of the detector to the bottom are primarily from short baselines of up to 80~km of atmosphere, while neutrinos that produce leptons that propagate directly up from the center of the earth were produced originally on the other side of the earth, some 10,000~km away.  

Accelerator-based neutrino beams allow for a much more intense neutrino beam that can be tuned to a specific energy and can be sent through the earth to produce measurements at a specific baseline (representing for each energy a specific time that the neutrino has had to evolve).  These neutrino sources start much like the atmospheric neutrino beam with protons hitting a target to produce pions and kaons, but in today's beams those pions and kaons are then focused by magnetic horns towards a decay pipe where they then decay to neutrinos whose directions are governed by the two-body kinematics of the pion decay.  The muons produced simultaneously with the neutrinos can also decay, which produce muon antineutrino and electron neutrino contamination.  The longer the decay pipe in the experiment, the more time the secondary muons have to decay to produce electron neutrinos.  The polarity of the focusing horn system can be changed to produce either a predominantly neutrino or antineutrino beam, at least at the focusing peak.  

Because there are comparable levels of neutrinos and antineutrinos in atmospheric neutrinos, and because the electron neutrinos make up almost a third of the neutrino beam, the atmospheric neutrino experiments are best suited for measuring muon neutrino disappearance over baselines of several orders of magnitudes, rather than making precision measurements of electron neutrino appearance.  The accelerator-based experiments are limited to a narrow range of distances to probe, but depending on the beamline focusing system a narrow or a wide neutrino energy spectrum can be produced in the neutrino beam.  

\begin{figure}[h]
\centering
\includegraphics[width=80mm]{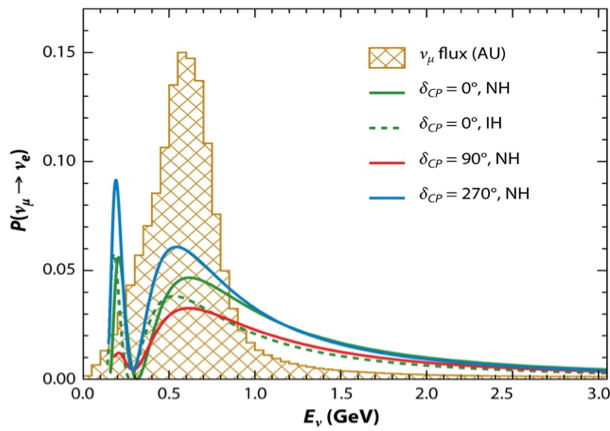}
\includegraphics[width=80mm]{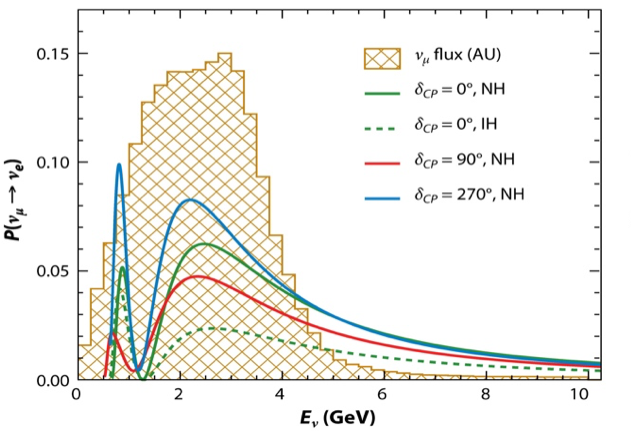}
\caption{The event spectra at Hyperk (top) and DUNE (bottom) overlaid with different possible oscillation probabilities.(taken from Ref.~\cite{milind}) } \label{spectra}
\end{figure}

The Hyper-K neutrino experiment will use a similar focusing system to the T2K experiment based at J-PARC which results in a narrow neutrino energy beam peaked at about 600~MeV.  The DUNE experiment will use a broad neutrino energy spectrum produced by the Long Baseline Neutrino Facility at Fermilab which ranges from 1 to 5~GeV.  The two spectra are shown in Fig.~\ref{spectra}, overlaid with oscillation probabilities which show how much of the oscillation process is covered by each experiment (taken from Ref.~\cite{milind}).  Typical electron neutrino contamination is on the order of a per cent or less at the focusing peak of neutrino energy, and the neutrino to antineutrino ratios at the focusing peak are also roughly 20 to 1, depending on the beamline geometry.  

\section{Neutrino Interactions} 

In order to measure the flavor and energy of a neutrino, oscillation experiments are limited to measuring (only some of) the outgoing particles created when a neutrino interacts in the detector.  For charged current neutrino interactions, the outgoing charged lepton flavor provides the incoming neutrino flavor, so measuring that lepton flavor and energy (and direction in the case of atmospheric neutrino experiments) is a minimum requirement.  

At energies that are comparable to the proton rest mass, the predominant neutrino interaction is quasi-elastic, where the incoming neutrino (antineutrino) interacts with a neutron (proton) in the detector which creates an outgoing charged lepton and proton (neutron).  If one were to assume that the initial nucleon were at rest, then the energy and direction of the outgoing lepton relative to the incoming neutrino would be enough to determine the energy of incoming neutrino through conservation of energy and momentum.  

However, most nucleons in today's neutrino detectors are inside nuclei. which means that even for this simple interaction the energy reconstruction is not so straightforward.  The initial nucleon is not at rest, and may also be correlated with other nucleons in the nucleus.  

An additional complication to reconstructing neutrino energies comes from the fact that for higher energy neutrinos, there are more interaction channels available, and the neutrino energy is spread between more final state particles.  This means that for an accurate energy estimate one would want to measure the energies of those final state particles, or at least have a good model of how the neutrino energy gets distributed among the various final state particles.  In addition, as those final state particles traverse the nucleus they too will be affected and lose energy and/or change direction.  

Given the statistical precision that the future neutrino facilities described here will reach, experiments will need to have accurate models of neutrino interactions in the nuclei of their detector materials to properly translate a measured neutrino energy spectrum into a "true"  neutrino energy spectrum.  These models are being developed now in the community and are aided by measurements of dedicated cross section experiments like MINERvA, as well as the cross section programs at current (T2K, NOvA) and future (SBND) near detector complexes.  

\section{Neutrino Detectors} 

The three neutrino detectors used in PINGU, Hyper-K and DUNE all have to identify muons and electrons and their energies at a minimum, but that is where the similarities end.  

PINGU is the largest of these three detectors, and as such has the lowest granularity, and highest neutrino energy thresholds.  PINGU consists of several strings of light sensors (phototubes) that are buried deep in ice at the South Pole, in the core of the km-scale ICECUBE detector.  The current DEEPCORE region of the ICECUBE detector has strings of optical modules that are separated by 42~m, and optical modules on the same string are separated by 7~m\cite{deepcore}.  For the future upgrade which will have yet a lower energy threshold, the phototube spacing on a single string will be 2.4~m, and the string separation will be 7~m instead of 7~m\cite{pingu}. In addition, PINGU is working on new optical sensors with higher performance.  For example, PINGU is considering modules that have two or several phototubes oriented in different directions on each optical module, and they are also investigating improved on-board calibration devices.  

The HyperKamiokande detector is a water-filled tank that measures 60~m tall and 74~m in diameter, lined with 50~cm diameter phototubes\cite{hyperk}.  This is an extension of the Super-Kamiokande water Cevenkov detector with almost twenty times the fiducial volume.  The number of phototubes planned will result in 40\% of the area of the walls, ceiling and floor of the tank to be active.  Improvements to the Hyper-Kamiokande design over that of Super-Kamiokande include optical modules that (also) contain more than one phototube per module for better light readout, better electronics to be ready for the high rate that a supernova burst would incur, and Gd doping.  If the Hyper-Kamiokande detector could be doped with Gd then it would have much higher efficiency for identifying antineutrino interactions, and could then be sensitive to reactor neutrino sources.  

The DUNE detector will consist of four modules of instrumented Liquid Argon.  The modules themselves have instrumented regions that measure 60~m x 12~m x 14~m.  The first module will use with planes of wires separated by 1.5 or 3~m, and each plane of wires has a wire pitch of 4.7~mm\cite{dune}.  
This technology has already been used in 10's of GeV energy neutrino beams (ICARUS\cite{icarus} in Gran Sasso) and in $~$1GeV neutrino beams (MicroBooNE\cite{microboone} at Fermilab), although again at an order of magnitude or more smaller scale than that of one of the DUNE far detector modules.  To prepare for the DUNE far detector scale, a smaller version of the detector was made with half-height but full width wire plane assemblies, and placed in a charged particle test beam at CERN in the fall of 2018.  That detector, called ProtoDUNE-SP (for Single Phase), saw both electron and pion beams that ranged in energy from 1~GeV to 10~GeV.

\begin{figure}[h]
\centering
\includegraphics[width=38.5mm]{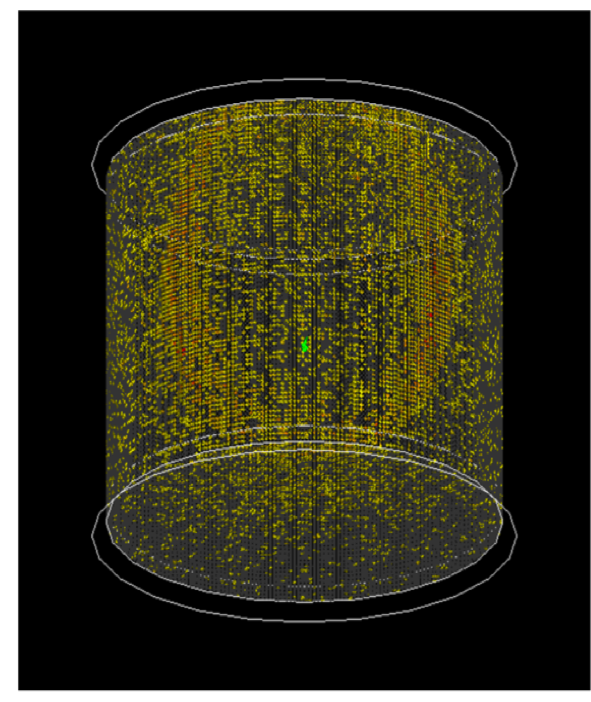}
\includegraphics[width=41.5mm]{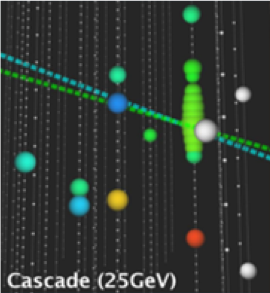}
\includegraphics[width=80mm]{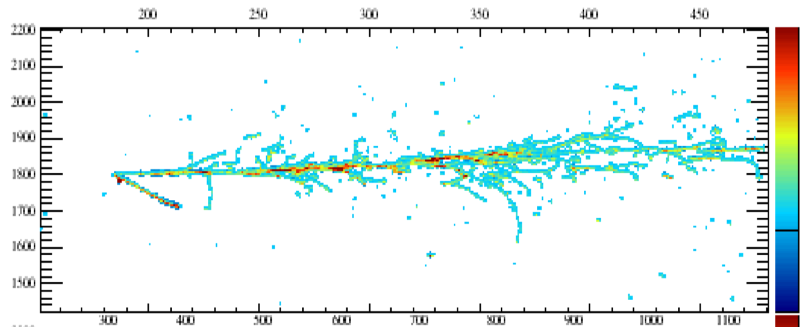}
\caption{The event displays for an electron neutrino charged current interactions at HyperKamiokande (top left), DeepCore (top right) and DUNE (bottom).} \label{evtdsp}
\end{figure}

Three event displays for these three detectors for a charged current electron neutrino interaction can be seen in Fig.~\ref{evtdsp}.  

\subsection{Cerenkov Light Detectors}

Cerenkov detectors work by collecting the light that is created in a medium when particles travel faster than the speed of light in that medium.    Both PINGU and Hyper-Kamiokande operate by this principle, albeit at very different energy regimes.  The minimum energy that Hyper-Kamiokande is sensitive to is a function of the noise on the phototubes and is on the order of a handful of MeV~\cite{hyperk}, while the minimum energy that PINGU is sensitive to is a function of the spacing of the optical modules.  The Cerenkov thresholds of different particles are determined by the index of refraction of the medium and the mass of the particles~\cite{pingu}.  The particles produced in PINGU tend to be well above Cerenkov threshold so in that case the detector measures the shower development (see Fig.~\ref{evtdsp} (top right)) and can see if a muon is produced by looking at a line of phototubes that saw light, where that line extends well past the shower region.  In Hyper-K, by contrast, the detector usually identifies final state leptons by the ring that is created by that single particle, as can be seen in Fig.~\ref{evtdsp}(top left).  Protons and pions created in Hyper-K's neutrino interactions tend to be below threshold, although when pions decay to muons and then electrons, the decay electrons often do show up as a delayed low energy electron-like ring and so can be identified in that way.  

\subsection{Time Projection Chambers} 

Time Projection Chambers (TPC's) consist of conductors (wires or pads) placed at regular intervals in a bath of liquid argon, and then a large electric field is induced which causes ionized electrons to drift towards the conductors which induces charge that is very localized.  The time of arrival of the electrons gives information about the distance perpendicular to the active plane, and then the wires or pads give spatial information in the other dimensions.  In this way TPC's can provide three-dimensional track reconstruction of charged particles, and the ionization along the track can be used for particle identification.  In the Fig.~\ref{evtdsp}(bottom) one can see clearly a short proton track and an electron shower originating from the same neutrino interaction vertex.  This is one of the simplest topologies, as the neutrino energy increases, more final state particles will be produced, and then show up as tracks in the liquid argon TPC.  

\section{Near Detectors} 

In order to reach the required precision on the oscillation measurements, experiments need precise predictions for what the visible energy spectrum will be for any set of oscillation probabilities.  That visible spectrum is not only a function of the oscillation probability, but also a convolution of the produced neutrino flux, the interaction probability, and the detector response.  
The {\em a priori} flux prediction comes from detailed neutrino beam simulations which much incorporate pion production cross sections which are themselves not perfectly known to better than 5-10 per cent.  In addition, the focusing geometry itself must include details that come from the focusing element geometry which itself may not be perfectly modeled. Finally, the neutrino interaction cross sections at the few-GeV energies of Hyper-K and DUNE are also not known.  

All of these uncertainties can be somewhat mitigated by the presence of one or more near detectors.  Those near detectors are also limited in that they can only measure a convolution of neutrino flux, interaction probability, and detector response, but by measuring the event rate with detectors with similar nuclei, and in part with similar detector response, the uncertainties in the underlying ingredients have a reduced effect on the final measurement.  

However, the challenge of near detector design comes from the fact that the flux at the near detector location is up to a factor of a million more intense, and the detector may not be under as much overburden as the far detector so the near detector may suffer from cosmic ray backgrounds in a way that the far detector does not. These factors are informing the current near detector designs of both DUNE and Hyper-Kamiokande.  

\subsection{Hyper-Kamiokande}

The Hyper-Kamiokande experiment, since it will be operating in the same neutrino beamline that T2K is currently operating, can take advantage of the T2K near detector complex.  That complex includes an on-axis detector and a suite of off-axis detectors that are in the path towards the Super-Kamiokande detector.  The on axis detector can see the neutrino event rate at the peak of the neutrino beam, but also it can see the fall-off of neutrino event rate as the detection point strays from the center of the beamline axis.  The current T2K off-axis near detector suite includes fine-grained scintillator detectors combined with gas TPC detectors which can do particle identification and charge identification.  There are also targets in the T2K near detector suite that allow for measurements on water instead of hydrocarbon.  

Plans are underway right now to build a new detector that will be in the path towards the Hyper-Kamiokande detector, and it will be able to measure the neutrino energy spectrum at a broad range of off-axis angles (Nu-PRISM)\cite{nuprism}.  By going to different angles the detector will see different fluxes, and by doing this suite of measurements the plan is to decouple the effects of flux and cross sections.  The challenge will be to understand the difference in detector efficiencies between the NuPRISM design which will use a 10~m wide instrumented cylinder instead of the  74~m diameter wide instrumented cylinder that is the Hyper-Kamiokande detector.  One idea to make the two detector images more similar is to use much smaller diameter phototubes on the Nu-PRISM detector, so that the granularity of the rings is more comparable between the two detectors, even if the actual diameter of the rings is very different.

\subsection{DUNE} 

The DUNE Near Detector Complex will feature a Liquid Argon TPC, but because of the high instantaneous neutrino flux combined with the slow drift times of electrons, the active elements will need to "see" much less argon than the far detector active elements.  One sobering statistic is that for a  $4 \times 3\times 5 m^3$  of vessel of liquid argon located about 1~km downstream of the neutrino target there will be 37 million  Charged Current $\nu_\mu$ interactions per year (assuming initial proton energy of 80~GeV and a yearly exposure of $1.5\times10^{21}$ protons on target)\cite{dune}.  

One way to solve this problem is to make the readout a plane of pixels rather than a plane of wires.  This results in a very high channel count for the near detector, which then requires very low power electronics so as not to heat up the Liquid Argon\cite{dwyer}.  But by having a several ton Argon near detector, the DUNE experiment can make interaction measurements on Argon specifically, which will mean that the nuclear effects on the various channels will be the same as in the far detector.  Again care must be taken to account for differences in the acceptance of the pixellated near detector and the wire-plane-based far detector.  There will also be other detector elements downstream of the Liquid Argon TPC to catch the outgoing hadronic showers and muons, in addition to extended capability to study neutrino-argon interactions in a gas TPC.  

DUNE is also planning to build in the capability to measure the neutrino flux at several different locations in order to make known specific changes to the neutrino flux and see how the resulting observed event energy distribution changes.   For the DUNE experiment this will be achieved by making a wide near detector hall, and then by moving the liquid Argon TPC (and associated downstream instrumentation) transverse horizontally to the direction of the neutrino beam.   To verify that the neutrino beamline is functioning as it should while the Liquid Argon detector is off-axis, there will also be a stationary on-axis plastic scintillator detector and spectrometer located on axis~\cite{dune-nd}.  

\begin{figure}[h]
\centering
\includegraphics[width=80mm]{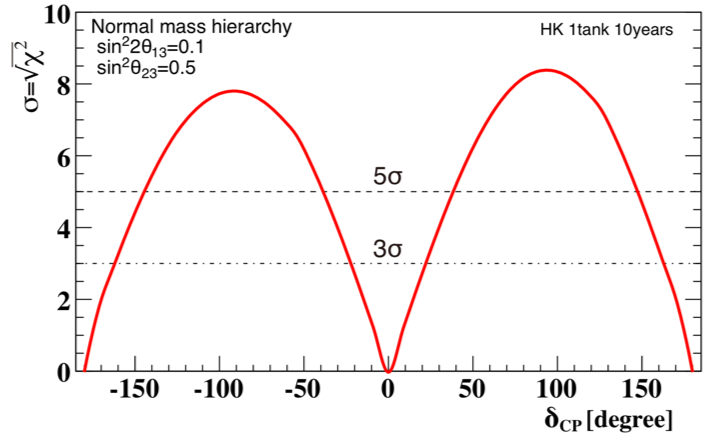}
\includegraphics[width=80mm]{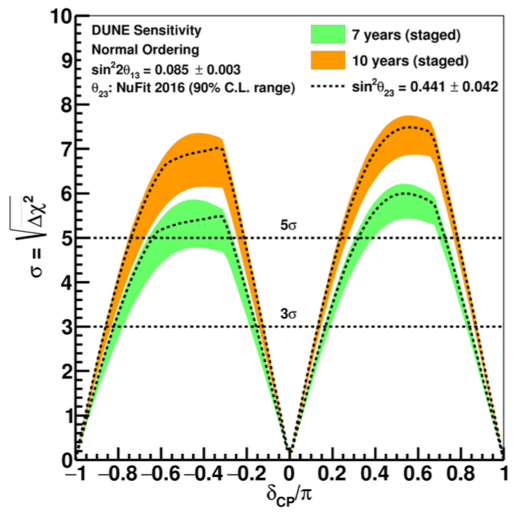}
\caption{The sensitivity of determining that CP violation is non-zero as a function of CP-violating phase $\delta$ for  HyperKamiokande (top, from Ref.~\cite{hyperk}) and DUNE (bottom, from Ref.~\cite{Grant:2017mwc}).} \label{cpv}
\end{figure}

\section{Physics Reach} 

DUNE and Hyper-Kamiokande are aiming to be able to see CP violation at over 5$\sigma$ over most of the phase space that $\delta$ can occupy.    Figure \ref{cpv} shows the number of sigma that CP violation can be seen at both Hyper-Kamiokande and DUNE, both after a 10-year run time at the projected beam intensities.  However, the important thing is that these reaches are achieved by extremely different strategies:  broad band versus narrow band neutrino beams, detectors with sensitivity to primarily the final state lepton that focus mostly on quasi-elastic events to detectors with sensitivity to a broad range of interaction channels.  The reach of Hyper-K is calculated assuming the mass ordering is known, while DUNE, because of its long baseline will be capable of determining the mass ordering in far less time than it will take to see CP-violation.  

\begin{figure}[h]
\centering
\includegraphics[width=80mm]{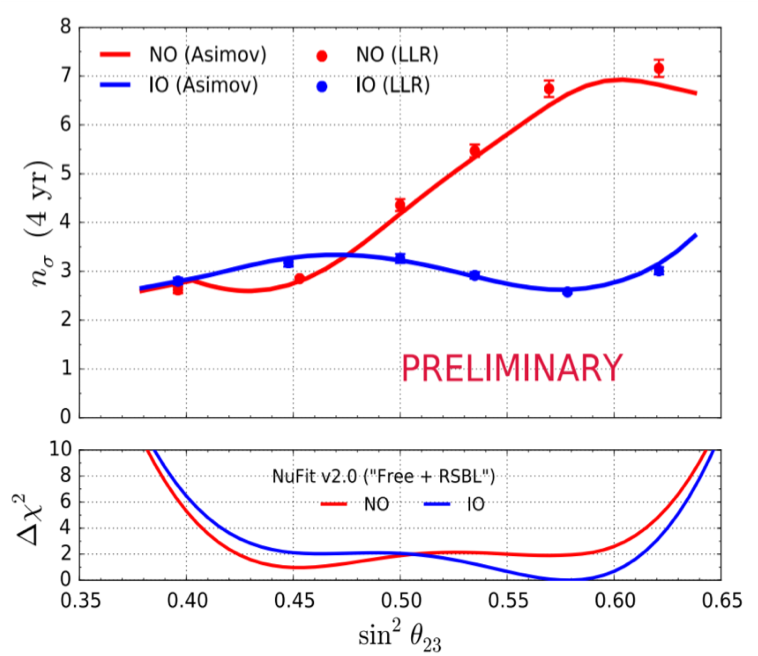}
\includegraphics[width=80mm]{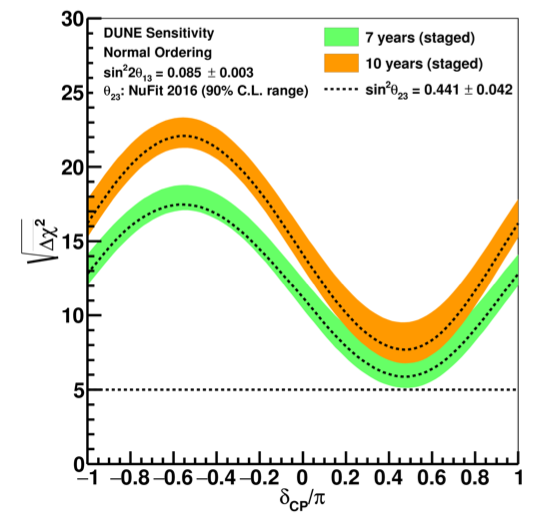}
\caption{The sensitivity of determining the mass ordering for (top) DUNE as a function of the CP-violating phase $\delta$  (from Ref.\cite{dune}) and for (bottom) PINGU (from Ref.~\cite{pingu}) as a function of the mixing angle $\theta_{23}$.} \label{mo}
\end{figure}

PINGU's primary sensitivity is to the mass hierarchy because PINGU can take advantage of the earth's matter effects.  Because of DUNE's long baseline, it is also sensitive to the mass ordering, and should be able to determine the neutrino mass ordering after only a few years of running.    Figure \ref{mo} shows the physics sensitivities to the mass ordering for both DUNE (after 7 or 10 years) and PINGU (after 4 years).  For DUNE the sensitivity to mass ordering depends a bit on the CP-violating phase $\delta$ while for PINGU because they will make a disappearance measurement the sensitivity has minimal dependence on the phase $\delta$.  

Again, the question of which experiment has better sensitivity to the mass ordering is not nearly as important as the fact that these very different experiments operating at energies that are factors of 3-10 apart, can access the same fundamental neutrino parameters.   Being able to compare one measurement with another where the sensitivities are comparable will allow for a host of new physics to be probed.  

\section{Conclusions} 
The particle physics community is now preparing three extremely different neutrino facilities whose goal is to elucidate flavor physics in the lepton sector, and to be prepared for surprises.  DUNE, HyperKamiokande, and PINGU will do this by probing neutrino energies from less than 1~GeV to dozens of GeV, by measuring neutrinos after they have passed through very little of the earth, and measuring neutrinos passing through the core of the earth.   These three facilities can make sure the framework that has been developed about three-flavor oscillations is valid.  
All three experiments also have significant non-flavor physics sensitivities as well:  for example, the three facilities would see neutrinos from a nearby supernovae in different ways.  Hyper-Kamiokande would be particularly sensitive to supernovae antineutrinos from interactions on the hydrogen in water, while DUNE would see supernovae neutrinos through the reaction of neutrinos to make $^{40}K^*$ from $^{40}Ar$.  On the way to these new physics measurements, there are also a host of opportunities to better understand neutrino interactions.  Regardless of what surprises await us, the field will be ready to investigate.  

\begin{acknowledgments}
This document was prepared using the resources of the Fermi National Accelerator Laboratory (Fermilab), a U.S. Department of Energy, Office of Science, HEP User Facility. Fermilab is managed by Fermi Research Alliance, LLC (FRA), acting under Contract No. DE-AC02-07CH11359.
The information in these proceedings and much of the graphics were contributed by Philip Eller and Masashi Yokoyama.  
The author wishes to thank Teresa Montaruli, Philip Eller, and Masashi Yokoyama for useful conversations in preparation of this contribution.  
\end{acknowledgments}

\bigskip 

\end{document}